\documentclass[12pt,a4paper]{article}
\usepackage{graphicx}
\usepackage{jcappub}
\usepackage{epsfig} 
\usepackage{amsmath}
\usepackage{amsfonts}
\usepackage{graphicx}
\usepackage{amssymb}

\title{On the Spin Bias of Satellite Galaxies in the Local Group-like Environment}
\author[a]{Jounghun Lee}%
\author[b]{and Gerard Lemson}%
\affiliation[a]{Astronomy Program, Department of Physics and Astronomy, Seoul National 
University, Seoul 151-747, Korea}
\affiliation[b]{Max-Planck-Institut f¨ur Astrophysik, Karl-Schwarzschild-Str. 1, 
85741 Garching b. M¨unchen, Germany}
\emailAdd{jounghun@astro.snu.ac.kr}

\abstract{We utilize the Millennium-II simulation databases to study the spin bias of dark subhalos in the Local 
Group-like systems which have two prominent satellites with comparable masses. Selecting the group-size 
halos with total mass similar to that of the Local Group (LG) from the friends-of-friends halo catalog and 
locating their subhalos from the substructure catalog, we determine the most massive (main) and second to the 
most massive (submain) ones among the subhalos hosted by each selected halo. When the dimensionless spin 
parameter ($\lambda$) of each subhalo is derived from its specific angular momentum and circular velocity at 
virial radius, a  signal of correlation is detected between the spin parameters of the subhalos and the 
main-to-submain mass ratios of their host halos at $z=0$: The higher main-to-submain mass ratio a host 
halo has, the higher mean spin parameter its subhalos have. It is also found that the correlations exist even 
for the subhalo progenitors at $z=0.5$ and $1$.  Our interpretation of this result is that the 
subhalo spin bias is not a transient effect but an intrinsic property of a LG-like system with higher main-to-
submain mass ratio, caused by stronger anisotropic stress in the region. 
A cosmological implication of our result  is also discussed. }


\begin{document}
\maketitle

\section{Introduction}\label{sec:intro}

The Local Group (LG) is a dumbbell-shaped group of galaxies which include the great Andromeda (M31),  
the Triangulum  (M33)  and the Magellanic Clouds (MC) as well as our home, the Milky Way (MW) 
\cite{LG}.   The two centers of the dumbbell shaped LG are nothing but MW and M31 which are known to 
have very similar masses of $\sim 10^{12}\,h^{-1}M_{\odot}$ \cite{MR,MWmass}.  
These two prominent galaxies contribute most of the total mass of LG,  $M_{\rm LG}$, which has been 
estimated to be $\log[M_{\rm LG}/M_{\odot}]=12.72$  with a $2\sigma$ range of $[12.26,\ 13.01]$  based on the 
accurate measurements of the distance and pair-wise speed between MW and M31 \cite{LW08}. 
The majority of the other LG member galaxies are the satellites of either MW or M31, having orders of magnitude 
lower masses.  It is expected that MW and M31 will eventually form a large central galaxy through a major 
merger between them when LG completes its virialization.
 
An intriguing question to ask is if and how the presence of two prominent galaxies in the LG and 
their mutual interaction caused their satellite galaxies to possess any biased or anomalous properties 
compared with other typical group galaxies.  Occurring rarely in the LG-like environment \cite{klimen-etal10}, 
the major merger event is one of those few mechanisms that can have a significant effect on the 
geometrical and physical properties of the subhalos.  For instance, ref.~\cite{knebe-etal11} claimed that 
the major mergers of the M31 progenitors should be responsible for the change of the satellite hosts between 
M31 and MW.  Ref.~\cite{fouquet-etal12} also attributed the detected vast polar structures of the MW dwarf 
satellites \cite{vpos} to the major mergers of the M31 progenitors \cite[see also,][]{ibata-etal13}.

Here, we  suggest a scenario that the spin parameters of the LG member galaxies have higher mean value 
than the that of the typical group galaxies due to the high anisotropic stress in the LG site.  
According to the recent study of ref.~\cite{libeskind-etal12}, the evolution of the angular momentum of a galaxy in 
the nonlinear regime is driven primarily by the local vorticity effect.  The dumbbell shape of LG and the ongoing 
gravitational interaction between its two prominent galaxies, MW and M31, reflects the enhanced anisotropic 
stress  in the local region around LG, which must have originated from the external tidal effect \cite{PC09}. 
Given that the spin parameter of a galactic halo is strongly correlated not only with its surface stellar density 
\cite[][and references therein]{KL12}  but also with the temperature and mass of its gas contents 
\cite{vasiliev-etal10},   understanding the spin parameter distributions of the LG member galaxies 
may provide a crucial key to explaining their physical properties as well as their evolution.
 
The goal of this Paper is to numerically test the above scenario by analyzing the data from 
the high-resolution N-body simulations. The contents of the upcoming sections are outlined as follows. 
In section \ref{sec:group} we describe the data from high-resolution N-body simulations and explain 
the numerical analysis of the simulation data. In section \ref{sec:bias} we present the main result on the 
spin bias of the dark subhalos in the LG-like systems and its redshift dependence. In section \ref{sec:con} 
we discuss a physical interpretation of our result and its cosmological implication as well.

\section{Main-to-submain mass ratios of group-size halos}
\label{sec:group}
\begin{figure}
\centering
\includegraphics[width=12cm]{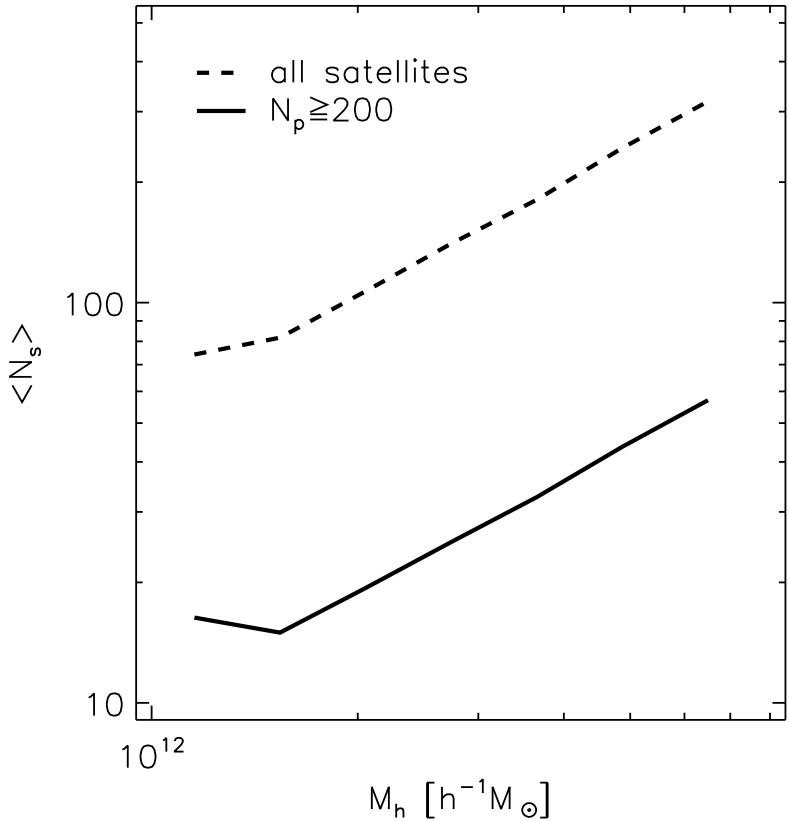}
\caption
{Mean numbers of the subhalos hosted by the group-size halos with mass $M_{h}$ 
at $z=0$. The solid and dashed lines represent the mean numbers of all subhalos 
and only those well resolved subhalos consisting of $200$ or more particles, respectively.}
\label{fig:pns}
\end{figure}
\begin{figure}
\centering
\includegraphics[width=12cm]{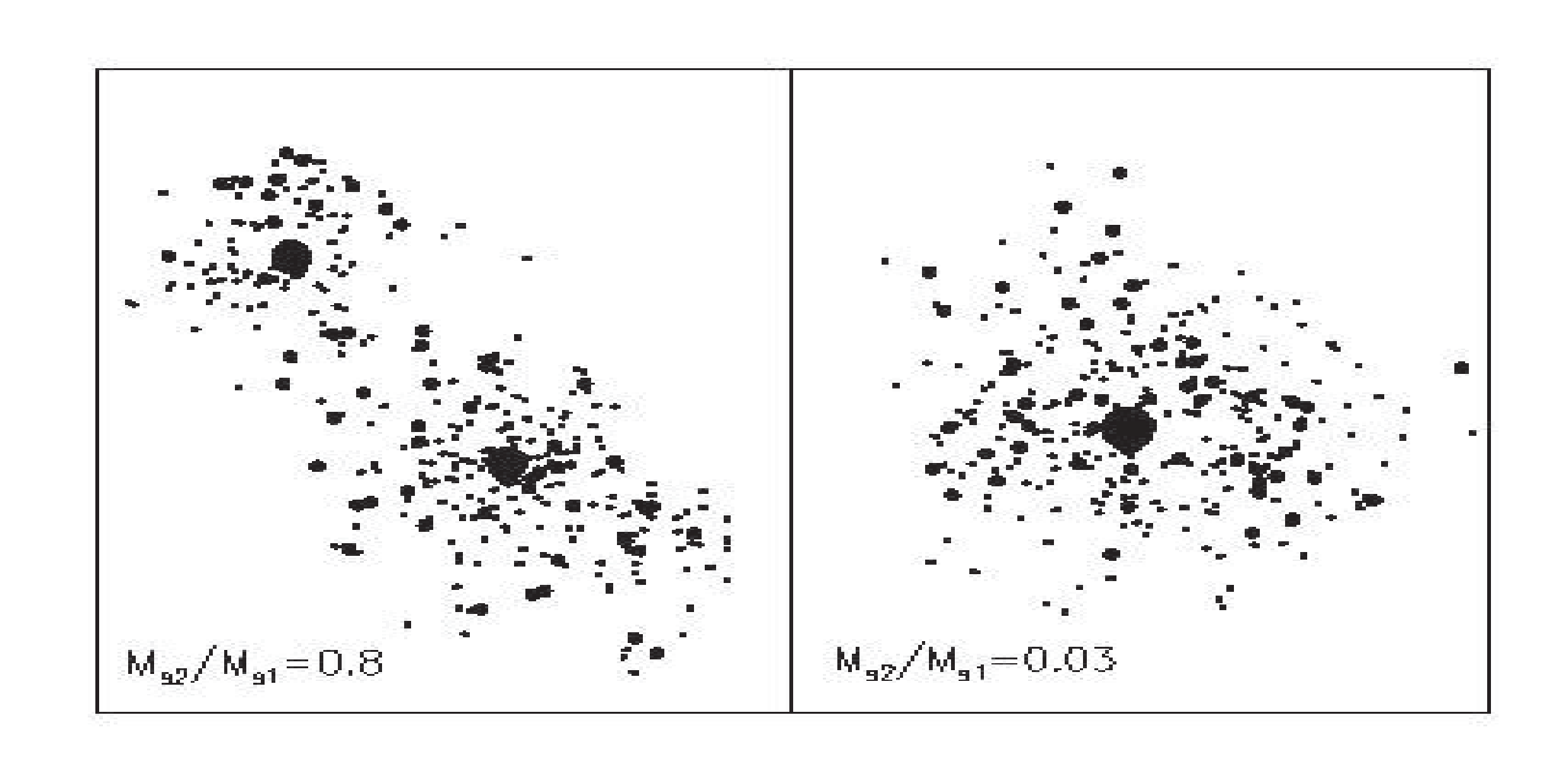}
\caption
{Spatial distributions of the subhalos in the two-dimensional projected space for the two 
different cases of the main-to-submain mass ratios at $z=0$. In each panel the medium-size 
filled circles and the small filled circles correspond to the well-resolved subhalos with 
$N_{p}\le 200$ and $N_{p}<200$, respectively. The two large filled circles in the left 
panel correspond to the main and submain subhalos while the one large filled 
circle in the right panel corresponds to the single main subhalo.}
\label{fig:proj}
\end{figure}
\begin{figure}
\centering
\includegraphics[width=12cm]{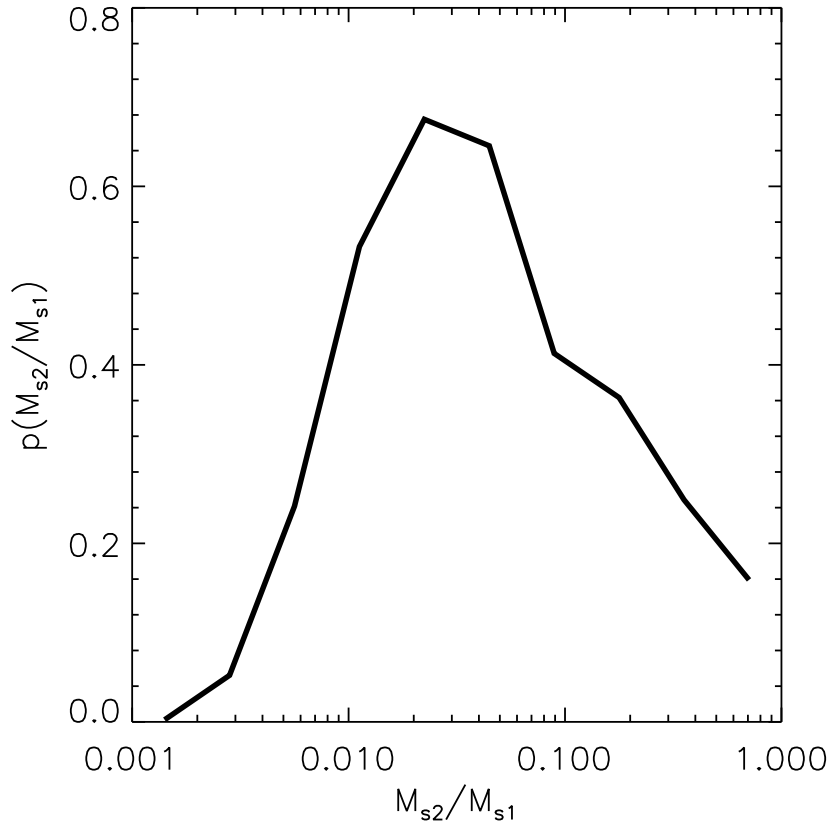}
\caption
{Probability density distribution of the main-to-submain mass ratios of the selected group-size 
halos at $z=0$.}
\label{fig:pr}
\end{figure}
\begin{figure}
\centering
\includegraphics[width=12cm]{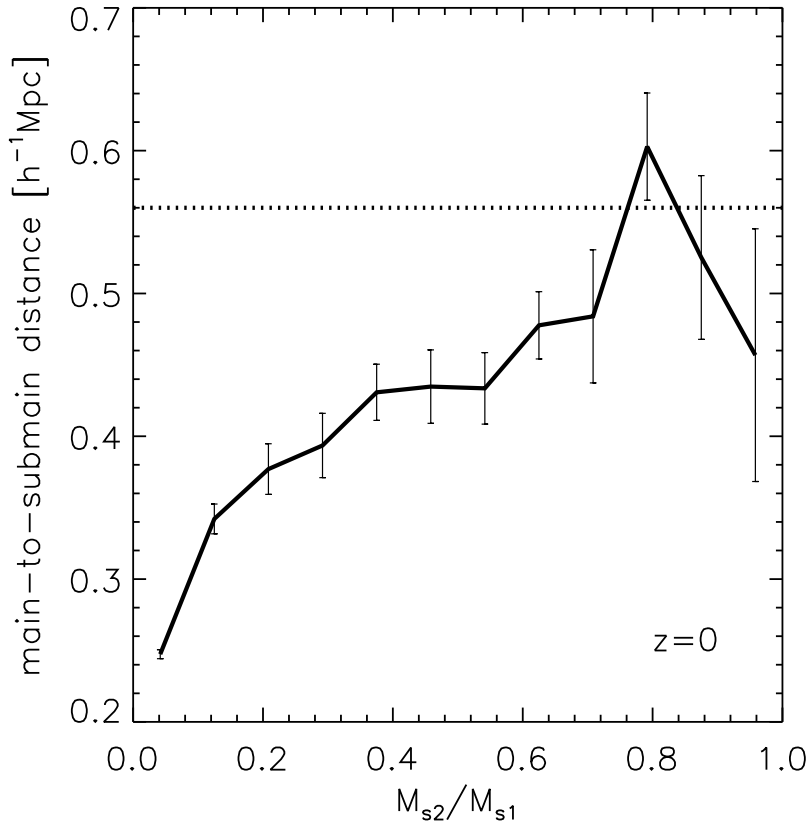}
\caption
{Separation distances between the main and the submain subhalos as a function of their mass ratios at $z=0$. 
The horizontal dotted line corresponds to the separation distance between the MW and the M31.}
\label{fig:distmr}
\end{figure}
\begin{figure}
\centering
\includegraphics[width=12cm]{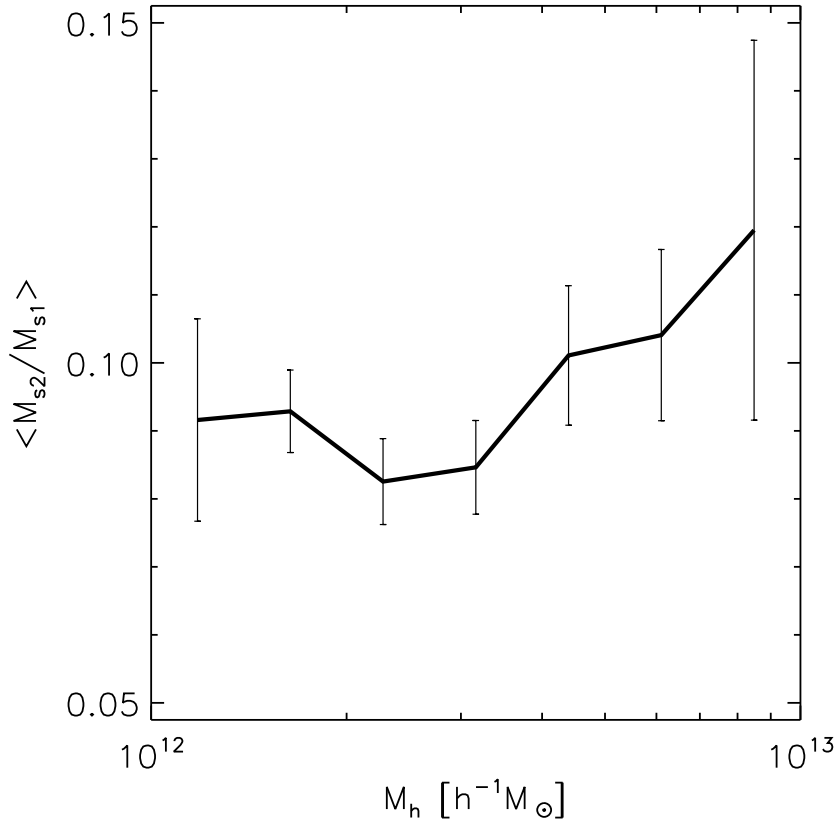}
\caption
{Mean main-to-submain mass ratios of the group-size halos as a function of their 
FoF mass at $z=0$.}
\label{fig:rmh}
\end{figure}
\begin{figure}
\centering
\includegraphics[width=12cm]{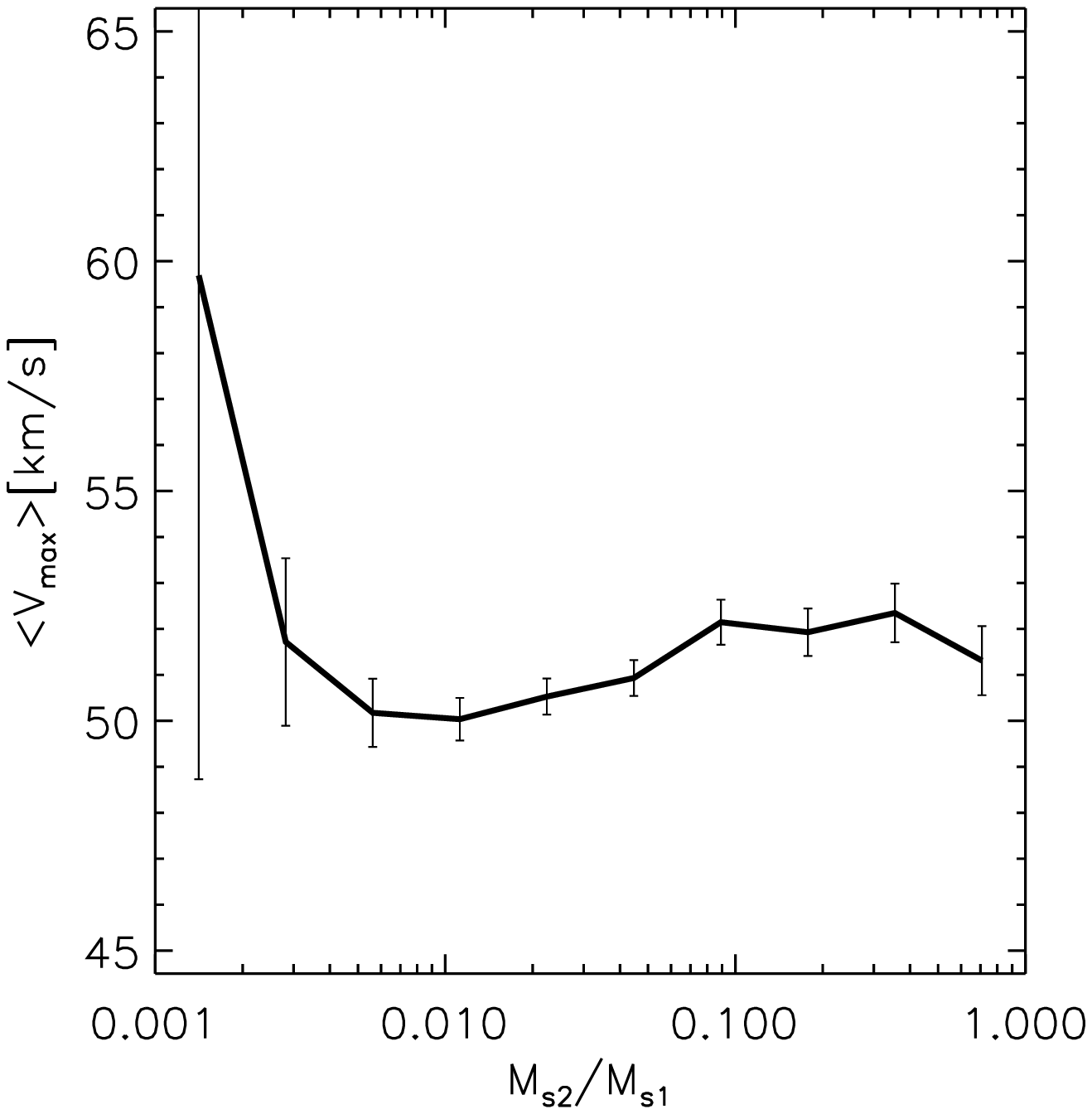}
\caption
{Mean circular velocity maximum of the subhalos as a function of their host halo's main-to-submain mass 
ratio at $z=0$.}
\label{fig:vmaxr}
\end{figure}

For the numerical investigation we utilize the catalogs of dark matter halos and their subhalos resolved 
in the Millennium II simulations \cite{millennium2}. Under the assumption of a flat $\Lambda$CDM 
cosmology with the key parameters of $\Omega_{m}=0.25,\ \Omega_{\Lambda}=0.75,\ n_{s}=1.0,\ 
\sigma_{8}=0.9,\ h=0.73$, the Millennium-II simulations were performed at various  epochs in a periodic box 
of linear size $100\,h^{-1}$Mpc with $2156^{3}$ dark matter particles each of which has mass of 
$6.89\times 10^{6}\,h^{-1}M_{\odot}$ . The dark matter halos and their subhalos were identified by 
applying the friends-of-friends (FoF) and the subfind algorithms \cite{subfind} to the particle data from the 
Millennium-II simulations, respectively.  The full descriptions of the Millennium-II simulation and how to 
retrieve information from the halo catalogs are provided in refs.~\cite{millennium2,lemson-etal06}, respectively. 

From the Millennium-II FoF catalogs at $z=0$, we first select those group-size halos whose FoF masses, 
$M_{h}$, are in the $2\sigma$ mass range of  LG \cite{LW08}: $10^{12.26}\le M_{h}/M_{\odot}\le 10^{13.01}$ . 
A total of $2079$ dark halos in the Millennium-II FoF catalog at $z=0$ are found to satisfy this mass constraint. 
For each selected group-size halo, we extract their subhalos from the Millennium-II subhalo catalog at $z=0$ 
but consider only those well-resolved ones consisting of $200$ or more dark matter particles ($N_{p}$) 
for our analysis. Figure \ref{fig:pns} plots the mean number of the well-resolved (all) subhalos versus the FoF 
masses of their host halos at $z=0$ as solid (dashed) line.  As can be seen, the mean numbers of the 
well-resolved subhalos with $N_{p}\ge 200$ increase monotonically with $M_{h}$ but do not exceed 
$100$ in the whole range of $M_{h}$.
  
We define the main and the submain subhalos of each halo as the most massive and the 
second to the most massive subhalos, respectively. Then, we assign each selected halo its unique 
value of the main-to-submain mass-ratio, $M_{s2}/M_{s1}$, where  $M_{s1}$ and $M_{s2}$ denote the 
masses of the main and submain subhalos, respectively.  If some halo has this mass-ratio close to unity, 
it is similar to the Local Group, having a dumbbell shape with two centers. Figure \ref{fig:proj} illustrates the 
spatial distributions of the subhalos  in the projected $x$-$y$ plane for the two different cases of 
$M_{s2}/M_{s1}$.  The left panel corresponds to the case that the main-to-submain mass ratio is close 
to unity with two prominent subhalos of comparable masses (largest filled circles). Note that most of the other 
subhalos for this case seem to be the satellites of these two prominent subhalos.  The right panel 
corresponds to the case where the main-to-submain mass ratio is much smaller than unity with one single 
central dominant subhalo. In each panel,  the medium and small-size filled circles represent the projected 
positions of the subhalos other than the prominent ones with $N_{p}\ge 200$ and $N_{p}<200$, respectively.
To see how rare the dumbbell-shaped systems are among the selected group-size halos, we bin the values of 
$M_{s2}/M_{s1}$ and count the numbers of the group-size halos belonging to each bin to determine the 
probability density distribution of $M_{s2}/M_{s1}$, the result of which is shown in Fig.~\ref{fig:pr}. As can be 
seen, the probability density reaches its maximum value around $M_{s2}/M_{s1}=0.02$, dropping rapidly as 
$M_{s2}/M_{s1}$ approaches unity.  

To see if the distances between the main and the submain subhalos depend on their mass-ratios, 
we also calculate the mean main-between-submain distances averaged over those  hosts belonging to each 
bin of $M_{s2}/M_{s1}$, the result of which is plotted in Fig.~\ref{fig:distmr}.  The horizontal dotted line 
corresponds to the separation distance between the MW and the M31 \cite{LW08}.  As can be seen, the mean 
distance between the main and the submain subhalos increases as $M_{s2}/M_{s1}$ increases. Note also 
that it matches the separation distance between MW and M31 when $M_{s2}/M_{s1}$ has the value around 
$0.8$, which indicates that those FoF halos with $M_{s2}/M_{s1}\ge 0.8$ are indeed similar to the LG.

To see if the main-to-submain mass-ratio of a host halo depends on its total mass, we bin the values of 
$M_{h}$ and calculate the mean value of $M_{s2}/M_{s1}$ averaged over those hosts belonging to each 
bin  of $M_{h}$. Figure \ref{fig:rmh} plots the mean value of the main-to-submain mass ratios of the selected 
group-size halos as a function of its FoF mass. The errors represent one standard deviation $\sigma_{r}$ 
in the measurement of $\langle M_{s2}/M_{s1}\rangle$ computed as $\sigma^{2}_{r}=
\left[\langle\left(M_{s2}/M_{s1}\right)^{2}\rangle-\langle M_{s2}/M_{s1}\rangle^{2}\right]/(N_{h}-1)$ 
where $N_{h}$ denotes the number of those group-size halos belonging to each bin of $M_{h}$.  
As can be seen in Figure \ref{fig:rmh}, the mean value,  $\langle M_{s2}/M_{s1}\rangle$, does not vary strongly 
with the total mass, $M_{h}$. 

To see if the mass distribution of the subhalos depends on the main-to-submain mass ratios of their 
host halos, we bin the values of $M_{s2}/M_{s1}$  and calculate the mean value of the maximum circular 
velocity, $V_{\rm max}$,  averaged over the subhalos whose host halos belong to each bin of 
$M_{s2}/M_{s1}$.  The Millennium-II substructure catalog provides information on $V_{\rm max}$ for 
each subhalo, which is a good indicator of the subhalo mass.
Figure \ref{fig:vmaxr} plots the average value of $V_{\rm max}$ as a function of the main-to-submain
mass ratios of their host halos.  The errors represent again one standard deviation in the measurements of 
$\langle V_{\rm max}\rangle$. As can be seen, there is only very weak, if any, correlation between $V_{max}$ 
and $M_{s2}/M_{s1}$ with the mean value of $V_{max}$ around $50\,s^{-1}km$, regardless of the value of 
$M_{s2}/M_{s1}$.

\section{Spin bias in the LG-like Environments}
\label{sec:bias}

\begin{figure}
\centering
\includegraphics[width=12cm]{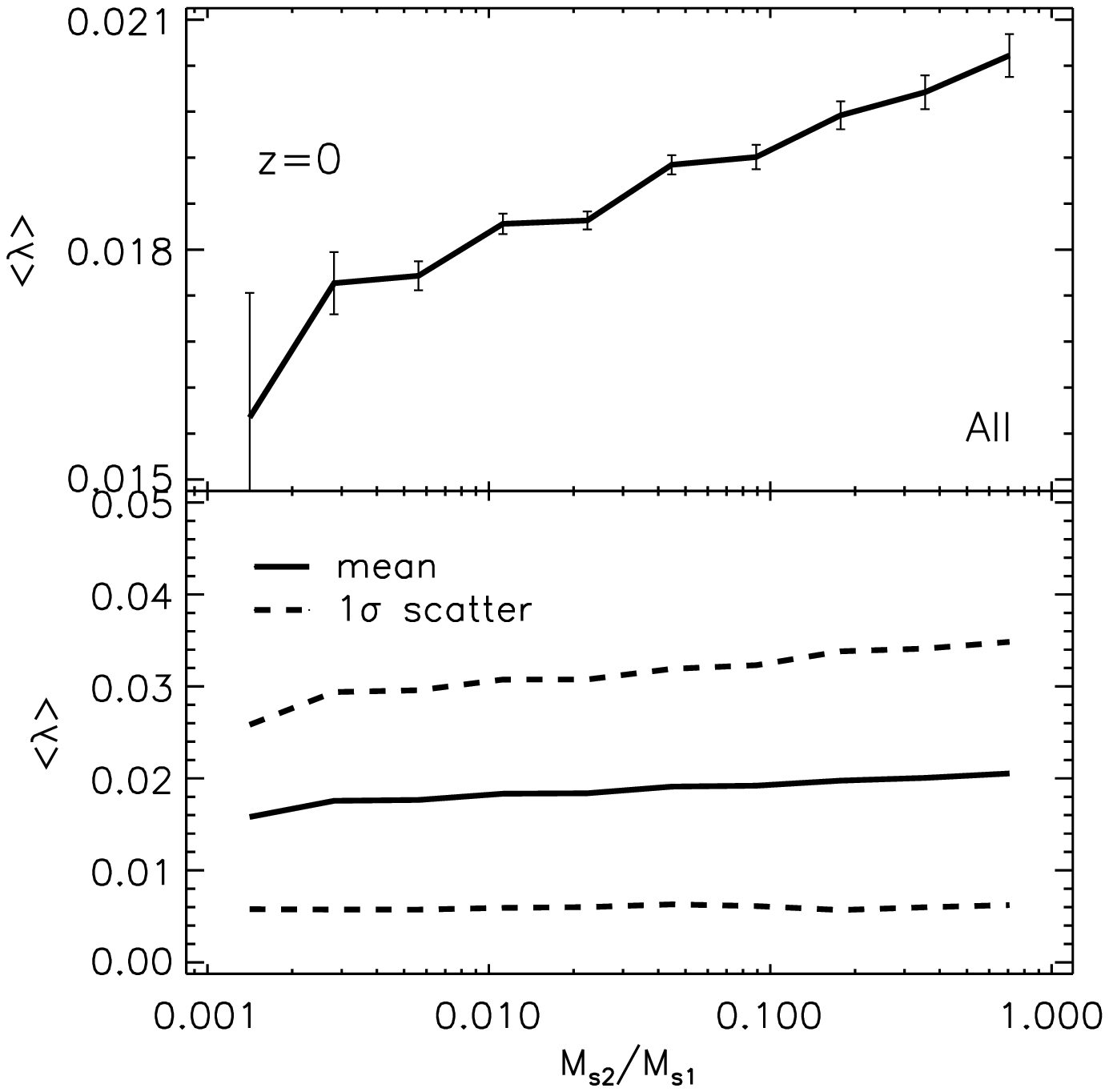}
\caption
{(Top panel): Mean spin parameter of the subhalos as a function of their host halo's main-to-submain mass 
ratio $M_{s2}/M_{s1}$ at $z=0$. (Bottom panel): Range of one standard deviation scatter 
(dashed line) of the spin parameters around its mean (solid line) vs. $M_{s2}/M_{s1}$.  }
\label{fig:lr}
\end{figure}

\begin{figure}
\centering
\includegraphics[width=12cm]{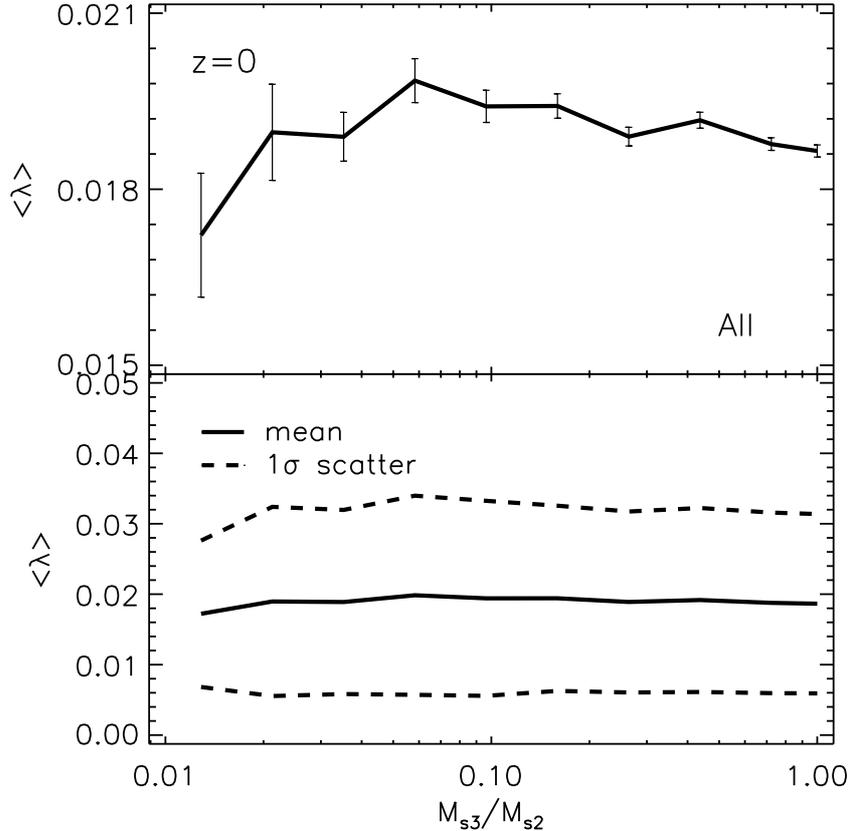}
\caption
{(Top panel): Mean spin parameter of the subhalos as a function of $M_{s2}/M_{s3}$ at $z=0$, where
$M_{s2}$ and $M_{s3}$ represent the masses of the second and the third to the most massive subhalos
belonging to a given host halo. (Bottom panel):  Range of one standard deviation scatter 
(dashed line) of the spin parameters around its mean (solid line) vs. $M_{s2}/M_{s3}$ . }
\label{fig:lr2}
\end{figure}

\begin{figure}
\centering
\includegraphics[width=12cm]{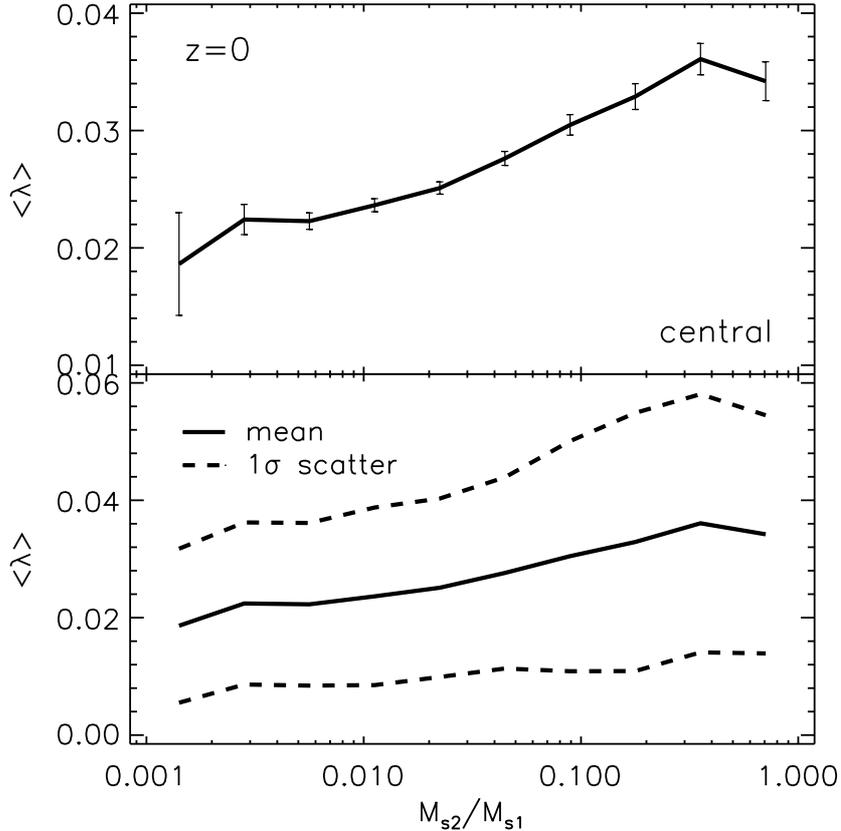}
\caption
{Same as Fig.~\ref{fig:lr} but only using the central prominent subhalos.}
\label{fig:lrc}
\end{figure}

\begin{figure}
\centering
\includegraphics[width=12cm]{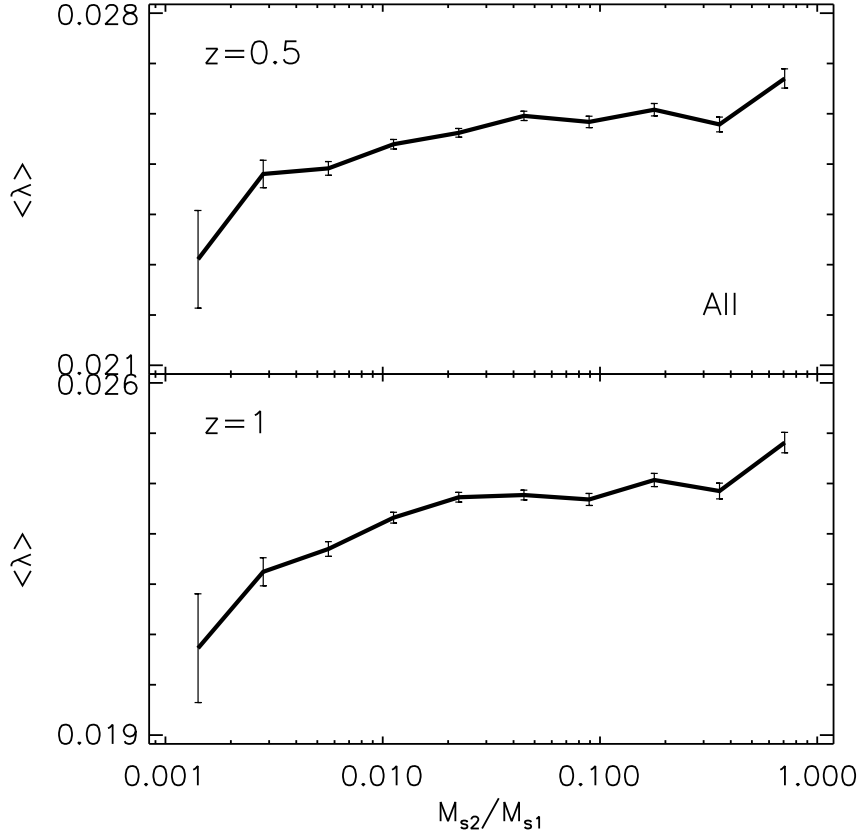}
\caption
{Correlations between the mean spin parameters of the subhalo progenitors and the main-to-submain 
mass ratios of the host halos at $z=0.5$ and $1$ in the top and bottom panels, respectively. .}
\label{fig:lrz}
\end{figure}

\begin{figure}
\centering
\includegraphics[width=12cm]{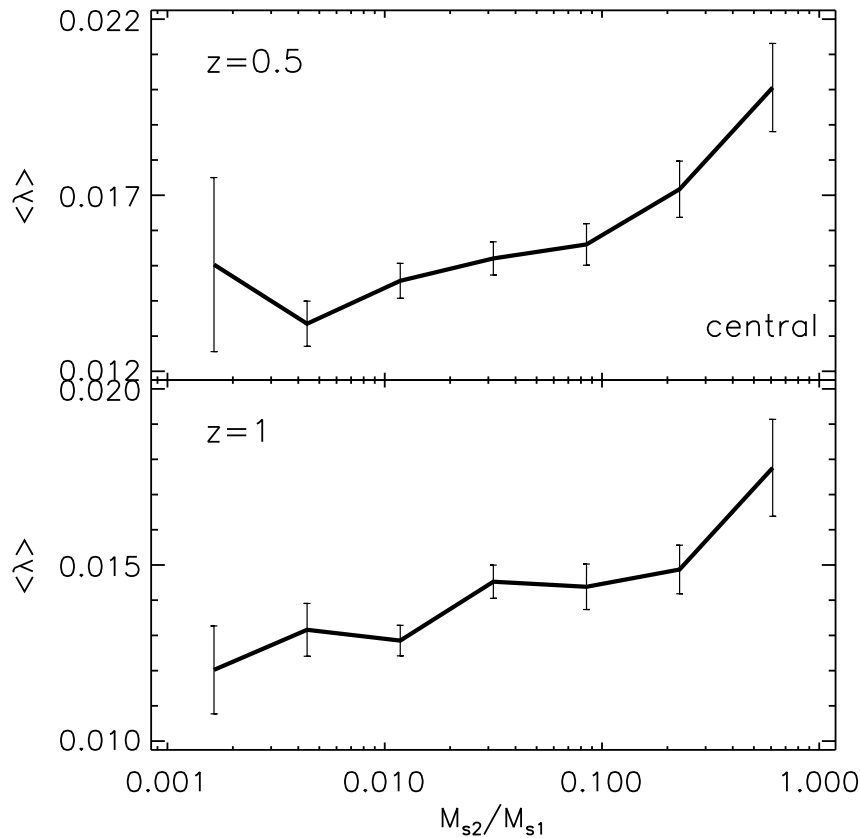}
\caption
{Same as Fig.~\ref{fig:lrz} but only for the central prominent satellites.}
\label{fig:lrzc}
\end{figure}

\begin{figure}
\centering
\includegraphics[width=12cm]{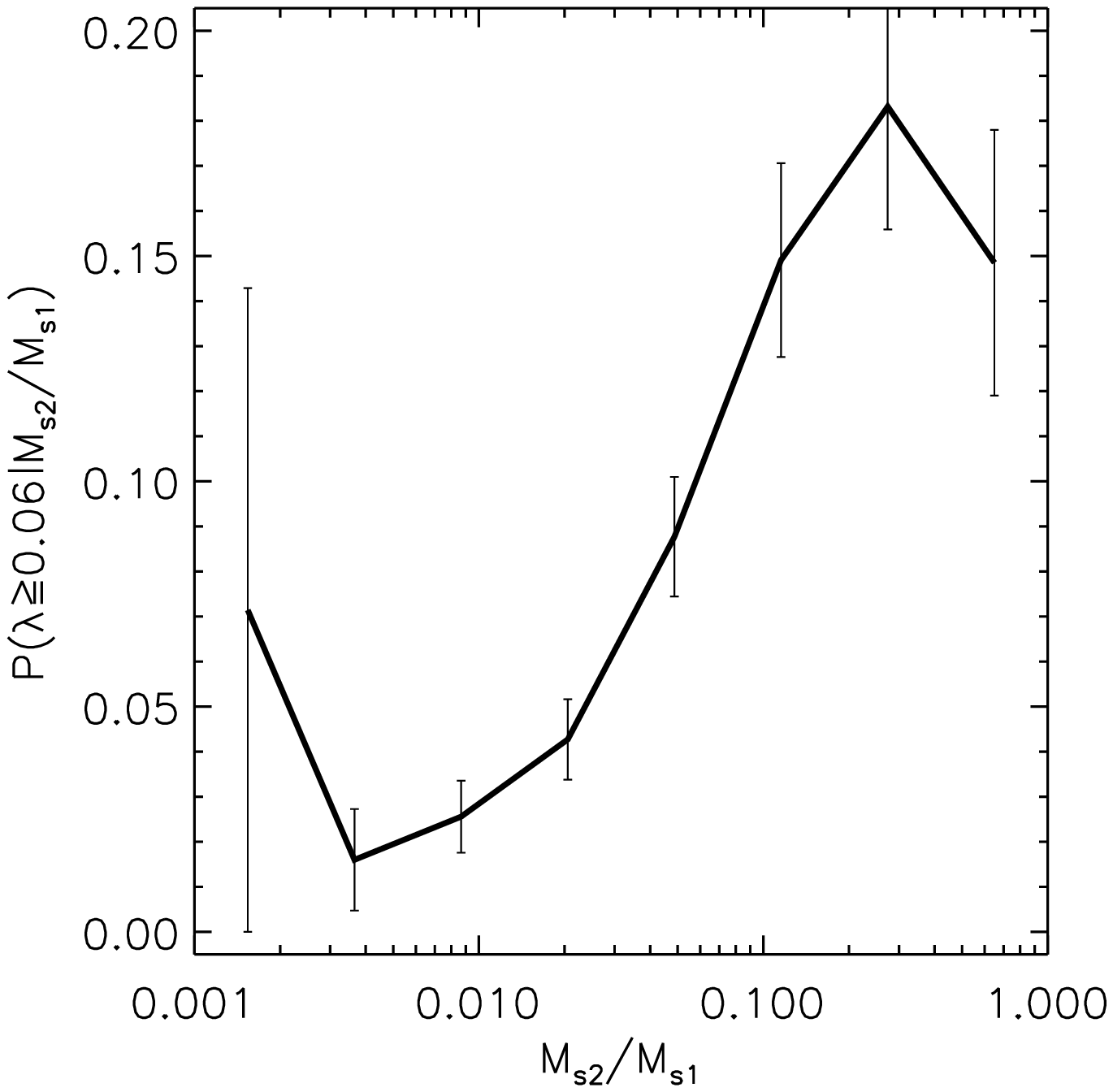}
\caption
{Number fractions of  the central subhalos with $\lambda\ge 0.06$ versus the main-to-submain 
ratio of the host halos at $z=0$. }
\label{fig:frac}
\end{figure}

Now that the LG-like groups are found atypical in the respect that most of the group-size halos with masses 
comparable to that of LG have main-to-submain halo ratios much less than unity, we would like to investigate
what environmental effect the atypical LG-like systems have on their subhalos.  We are particularly interested 
in the environmental effect on the subhalo's dimensionless spin parameter $\lambda$ which is conveniently 
defined as $\lambda = j/(\sqrt{2}V_{\rm vir}R_{\rm vir})$ \cite{bullock-etal01} where $R_{\rm vir}$ 
represents the virial radius and $V_{\rm vir}$ is the circular velocity measured at $R_{\rm vir}$, $j$ 
 is the magnitude of the subhalo's specific angular momentum (angular momentum per mass). 
For the subhalos identified by the SUBFIND algorithm, the virial radius $R_{\rm vir}$ is related to the 
spherical radius $R_{\rm max}$ at which the subhalo's circular velocity curve reaches its maximum as 
$R_{\rm vir}=R_{\rm max}/0.18$ \cite{muldrew-etal11}, while $V_{\rm vir}$, can be calculated 
from the virial mass $M_{\rm vir}$ and the virial radius $R_{\rm vir}$ as 
$V_{\rm vir}=\sqrt{GM_{\rm vir}/R_{\rm vir}}$ where $G$ is the Newtonian constant. 
Using these relations along with information on $j$, $V_{\rm max}$ and $R_{\rm max}$ provided in the 
Millennium-II substructure catalog, we compute the dimensionless spin parameter $\lambda$ of each 
selected subhalo.

Binning the mass ratio $M_{s2}/M_{s1}$ of each selected group-size halo and calculating the mean value of the 
spin parameters of those well-resolved subhalos whose host halos belong to each bin of $M_{s2}/M_{s1}$, we 
determine $\langle\lambda\rangle$ as a function of $M_{s2}/M_{s1}$, which is shown in the top panel of
Fig.~\ref{fig:lr}. The errors represent one standard deviation $\sigma_{\lambda}$ 
in the measurement of $\langle\lambda\rangle$ computed as $\sigma^{2}_{\lambda}=
\left[\langle\lambda^{2}\rangle-\langle\lambda\rangle^{2}\right]/(N_{h}-1)$ 
where $N_{h}$ denotes the number of those group-size halos belonging to each bin of $M_{s2}/M_{s1}$.
As can be seen, there exists a clear signal of correlation between $\lambda$ and 
$M_{s2}/M_{s1}$: The higher main-to-submain mass ratio a host halo has, the higher mean spin parameters 
their subhalos have. Recalling that the main-to-submain mass ratio of a host halo has no mass bias (see
Fig.~\ref{fig:vmaxr}) and that the subhalo spin parameters are insensitive to the subhalos' mass,  
we affirm that the correlation detected between $\lambda$ and $M_{s2}/M_{s1}$ is not due to any 
mass bias.  

The bottom panel of Fig.~\ref{fig:lr} shows one standard deviation scatter of $\lambda$ (dotted line), computed as 
$\left[\langle\lambda^{2}\rangle-\langle\lambda\rangle^{2}\right]^{1/2}$,  around its mean value (solid line). 
Although the width of the scatter of $\lambda$ is much wider than the range of the detected trend in $\lambda$ 
with $M_{s2}/M_{s1}$, it does not necessarily mean that the correlation between $\lambda$ and 
$M_{s2}/M_{s1}$is not meaningful since the spin parameter $\lambda$ is well known to be widely 
scattered following the log-normal distribution \cite{bullock-etal01}.   The presence of the correlation between 
$\lambda$ and $M_{s2}/M_{s1}$ is important and meaningful because it implies that for the case of higher 
$M_{s2}/M_{s1}$ the fraction of $\lambda\ge\lambda_{c}$ in the log-normal tail will be larger where 
$\lambda_{c}$ is some threshold of the spin parameter.

Now that a signal of correlation between $\lambda$ and $M_{s2}/M_{s1}$ is detected, it is interesting 
to examine whether or not $\lambda$ also depends on $M_{s3}/M_{s2}$ where $M_{s3}$ denotes the mass of the 
third to the most massive subhalo.  We repeat the same calculation to determine $\langle\lambda\rangle$ but 
as a function of $M_{s3}/M_{s2}$, the result of which is shown in Fig.~\ref{fig:lr2}.  As can be seen, the mean 
spin parameter of the subhalos depends weakly on $M_{s3}/M_{s2}$, reaching the maximum value at 
$M_{s3}/M_{s2}\approx 0.05$ and decreasing as $M_{s3}/M_{s2}$ increases. This result implies that the LG may 
be the optimal environment for the highest spin parameters:  
In addition to its high value of $M_{s2}/M_{s1}\approx 0.8$ \cite{MR},  the value of $M_{s3}/M_{s2}$ of the 
LG is approximately $0.05$ since the Triangulum galaxy (the third to the most massive member galaxy in 
the LG)  has mass approximately  $M_{s3} = 5\times 10^{10}\,h^{-1}M_{\odot}$ \cite{M33}.

Since it is only the central galaxies whose physical properties are known to depend on the spin 
parameters of their host halos \cite{jimenez-etal97,jimenez-etal98,mmw98}, we repeat the whole calculations using 
only the central prominent subhalos, the result of which is shown in Fig.~\ref{fig:lrc}. As can be seen,  we observe 
stronger correlation between the spin parameters of the central prominent subhalos and the main-to-submain mass ratios 
of their host halos. From this results, it can be inferred that the subsequent tidal stripping effect tends to reduce 
the strength of the correlation between the spin parameters of their subhalos and the main-to-submain mass ratios of their 
host halos.

Given that those host halos with higher $M_{s2}/M_{s1}$ are likely to be recent merger remnants, it should 
be worth checking whether or not the observed correlation between $\lambda$ and $M_{s1}/M_{s2}$ is a 
transient effect. Locating the progenitors of the subhalos belonging to each host halo at higher redshifts, $z=0.5$ 
and $z=1$, in the Millennium Merger Tree catalog, we investigate the correlations between the spin parameters of the 
subhalo progenitors and the main-to-submain mass ratios of their descendant hosts at $z=0$. Figure \ref{fig:lrz} shows 
the same as Fig.~\ref{fig:lr} but for the subhalo progenitors at $z=0.5$ and $z=1$ in the top and bottom panels, 
respectively. The results are obtained by considering only those well resolved subhalo progenitors with 
$N_{\rm p}\ge 200$. As can be seen, the mean spin parameters of the subhalo progenitors are correlated with 
the main-to-submain mass ratios of the descendant hosts and the strength of the correlations are similar 
to that observed at $z=0$. Using only the central prominent subhalos,  we repeat the whole calculations, the 
result of which is shown in Fig.~\ref{fig:lrzc}. As can be  seen,  we observe stronger correlation between the spin 
parameters of the progenitors of the central prominent subhalos and the main-to-submain mass ratios of their 
descendant hosts.  Noting the results shown in Fig.~\ref{fig:lrz}, we think that the observed trend in 
$\lambda$ with $M_{s2}/M_{s1}$ is not a mere transient phenomena due to the merging 
but an intrinsic effect of the anisotropic stress  which increases with the main-to-submain mass ratios.

Previous theoretical studies asserted that the surface stellar density of a disc galaxy is inversely proportional to 
the spin parameter of its host halo and that a dark galaxy whose surface stellar density falls below 
$100\,{\rm pc}^{-2}$ forms in the critically fast spinning halos whose spin parameters exceeds some 
threshold of $\lambda_{c}\approx 0.06$ \citep{jimenez-etal97,jimenez-etal98,mmw98}.  Very recently, 
ref.~\cite{KL12} performed a high-resolution hydrodynamic simulation to numerically confirm that 
the spin parameters are indeed strongly correlated with the surface stellar and gas densities of a disc galaxy. 
Their result revealed clearly that the halo's higher spin leads to the lower stellar and gas surface densities 
of its disk galaxy.  
To see how abundant the dark galaxies are in the LG-like systems, we calculate the number fraction of the 
prominent subhalos with $\lambda\ge \lambda_{c}=0.06$ as a function of the main-to-submain mass ratio, 
the result of which is plotted in Fig.~\ref{fig:frac}.  As can be seen, the fraction of  the fast-spinning central 
subhalos with $\lambda\ge\lambda_{c}$ increases almost monotonically with the main-to-submain mass ratio.  
For the case of $M_{s2}/M_{s1}\ge 0.3$, approximately $18\%$ of the central subhalos have 
$\lambda\ge\lambda_{c}$ while for the case of $M_{s2}/M_{s1}\le 0.05$ only $2\%$ of the central 
subhalos satisfy the condition.  

\section{Summary and discussion}\label{sec:con}
 
By analyzing the halo and subhalo catalogs from the Millennium-II simulations, we have detected a clear 
signal of correlations between the main-to-submain mass ratios of group-size halos and the spin 
parameters of their subhalos at present epoch. We have also found that the central prominent subhalos 
exhibit stronger correlations and that the correlations with similar strength exist even for the subhalo progenitors 
at $z=0.5$ and $z=1$.  We conclude that the observed high mean spin parameters of the subhalos 
in the LG-like groups are not transient merger remnants but likely to be intrinsic property of the LG-like systems 
induced by the high anisotropic stress in the local site. 

An important implication of our result is that the LG member galaxies are biased toward the high spins and thus 
likely to have on average lower stellar and gas surface densities than the typical group galaxies 
\cite{jimenez-etal97,jimenez-etal98,mmw98,KL12}. 
When the observed properties of the MW satellites  are used to be compared with the predictions 
of $\Lambda$CDM cosmology the spin bias of the MW satellites should be taken into account. For instance, our 
result may help alleviate the tension between the observed satellite populations of the MW and the predictions of 
the $\Lambda$CDM cosmology \cite[][and references therein]{missing1,missing2} since the presence of 
more dark satellite galaxies due to their biased spins in the MW system could explain the lower abundance 
of the observed satellites of the MW.
It is, however, worth mentioning here that the robustness of our result obtained from the Millennium-II  
simulations against the subhalo-finding algorithm will have to be tested further in the future before connecting 
it to the observed properties of galaxies.  Especially the masses and spin parameters of the subhalos calculated 
using the number of DM particles can have large variations in their values at the consecutive time steps due to 
the limitation of the SUBFIND algorithm. 

\acknowledgments 

We thank an anonymous referee for providing many useful comments which help us improve the original 
manuscript. 
The Millennium-II Simulation databases used in this paper and the web application providing online 
access to them (http://www.mpa-garching.mpg.de/galform/millennium-II/) were constructed as part of 
the activities of the German Astrophysical Virtual Observatory. The work of JL was supported by the 
National Research Foundation of Korea (NRF) grant funded by the Korea government (MEST, 
No.2012-0004916) and partially by the research grant from the National Research Foundation of 
Korea to the Center for Galaxy Evolution Research  (NO. 2010-0027910). The work of GL was 
supported by Advanced Grant 246797 "GALFORMOD" from the European Research Council.

\end{document}